\documentclass[%
 aip,
 jmp,%
 amsmath,amssymb,
 reprint,%
]{revtex4-1}
\usepackage{hyperref}
\usepackage{graphicx}
\usepackage{dcolumn}
\usepackage{bm}
\usepackage{color}
\begin{document}

\preprint{AIP/123-QED}

\title{High mobility dry-transferred CVD bilayer graphene}

\author{Michael Schmitz}
 \affiliation{2nd Institute of Physics and JARA-FIT, RWTH Aachen University, 52074 Aachen, Germany}
\author{Stephan Engels}%
\affiliation{2nd Institute of Physics and JARA-FIT, RWTH Aachen University, 52074 Aachen, Germany}
\affiliation{Peter Gr\"{u}nberg Institute (PGI-9), Forschungszentrum J\"{u}lich, 52425 J\"{u}lich, Germany}
\author{Luca Banszerus}
\affiliation{2nd Institute of Physics and JARA-FIT, RWTH Aachen University, 52074 Aachen, Germany}
\author{Kenji Watanabe}
\affiliation{National Institute for Materials Science, 1-1 Namiki, Tsukuba 305-0044, Japan}
\author{Takashi Taniguchi}
\affiliation{National Institute for Materials Science, 1-1 Namiki, Tsukuba 305-0044, Japan}
\author{Christoph Stampfer}
\affiliation{2nd Institute of Physics and JARA-FIT, RWTH Aachen University, 52074 Aachen, Germany}
\affiliation{Peter Gr\"{u}nberg Institute (PGI-9), Forschungszentrum J\"{u}lich, 52425 J\"{u}lich, Germany}
\author{Bernd Beschoten}
\affiliation{2nd Institute of Physics and JARA-FIT, RWTH Aachen University, 52074 Aachen, Germany}

\date{\today}

\begin{abstract}
We report on the fabrication and characterization of high-quality chemical vapor-deposited (CVD) bilayer graphene (BLG). In particular, we demonstrate that CVD-grown BLG can be detached mechanically from the copper foil by an hexagonal boron nitride (hBN) crystal after oxidation of the copper-to-BLG interface. Confocal Raman spectroscopy reveals an AB-stacking order of the BLG crystals and a high structural quality. From transport measurements on fully encapsulated hBN/BLG/hBN Hall bar devices we extract charge carrier mobilities up to 180,000~cm$^2$/(Vs) at 2~K and up to 40,000~cm$^2$/(Vs) at 300~K, outperforming state-of-the-art CVD bilayer graphene devices. Moreover, we show an on-off ratio of more than 10,000 and a band gap opening with values of up to 15 meV for a displacement field of 0.2 V/nm in such CVD grown BLG.
\end{abstract}

\maketitle

Bilayer graphene (BLG) is a unique two-dimensional material hosting massive chiral quasi-particles,\cite{McCann2013}
which give rise to
a number of interesting 
phenomena. This includes, for example, an unconventional quantum Hall effect,\cite{Novoselov2006} a tunable Lifshitz transition\cite{Varlet2014} or topologically protected valley transport.\cite{Shimazaki2015,Sui2015} Most importantly, BLG offers high carrier mobility and the possibility of opening a tunable band gap by breaking the intrinsic inversion symmetry using a perpendicular electric field,\cite{McCann2006,Min2007,Oostinga2008} (the so-called displacement field) making this material particularly interesting for electronic applications.

While most of the 
experimental work has been employed by using mechanical exfoliation of BLG from natural graphite, there are only a few reports on implementing synthetically grown BLG into high performance devices.\cite{Hao2016,Liu2012}
Chemical vapor deposition (CVD) is among the most promising growth methods when aiming for large-scale applications. The rapid advancement in this field has recently lead to the growth of continuous single-layer graphene (SLG) sheets as well as individual flakes of arbitrary lateral sizes ranging up to centimeter scale.\cite{Hao2013, Li2011, Chen2013, Zhou2013} For years, the electronic properties of the CVD-grown SLG and BLG could hardly compete with high-quality mechanically exfoliated samples.\cite{Petrone2012, Wang2013} The main reasons for the striking difference in device performance are material degradations and contaminations caused by wet transfer techniques which typically have been used for separating CVD graphene from the catalyzing metal surface. \cite{Suk2011} The direct exposition of graphene to wet chemicals such as etchants and solvents introduce defects, ripples and inhomogene doping in graphene and thereby results in minor electronic performance when incorporated into transport devices.\cite{Suk2011} Recently, we introduced a dry transfer technique where van-der-Waals forces are used to directly pick-up CVD-grown SLG from the metal surface by hexagonal boron-nitride (hBN) stamps.\cite{Banszerus2015, Banszerus2016} Using this method the SLG exhibit a high structural quality and ultra-high carrier mobilities up to $3\times 10^6 \text{cm}^2\text{/(Vs)}$.
This dry transfer method has not yet been demonstrated for CVD-grown BLG.

In this Letter, we present the fabrication of high-quality CVD-grown BLG fully encapsulated in hBN with charge carrier mobilities matching the largest values reported so far for exfoliated BLG at room temperature.\cite{Dean2010} Several hBN/BLG/hBN heterostructures are fabricated by a dry transfer method where the BLG crystals are mechanically detached from the underlying Cu-foil. Confocal Raman spectroscopy on encapsulated structures proves the AB-stacking order of the CVD-grown BLG and demonstrates its high structural quality in terms of low
strain inhomogeneities. Low temperature transport measurements of a top-gated Hall bar device reveal ballistic transport thanks to its high charge carrier mobility and allow exploring the formation of the band gap when applying a perpendicular displacement field.

\begin{figure}
	\centering
	\includegraphics[width=0.97\linewidth]{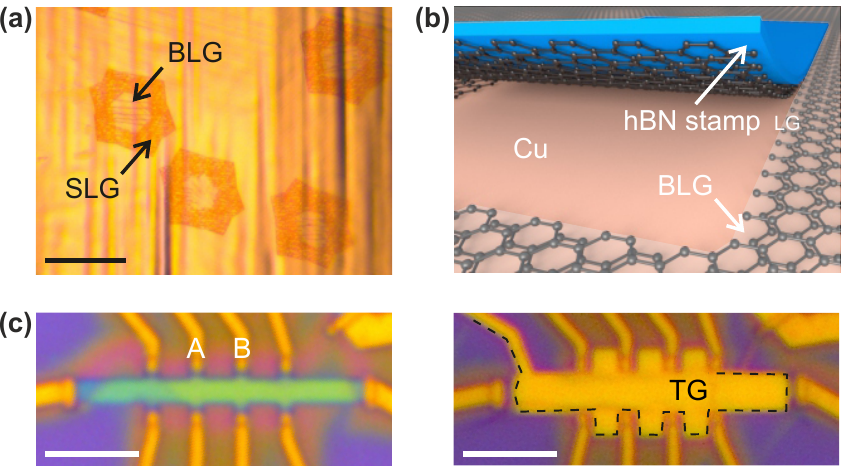}
	\caption{(a) Optical image of CVD-grown SLG crystals on a Cu foil with a BLG area in their centers. The scale bar represents 100~$\mu$m. (b) Schematic of mechanical delamination process. (c) Optical image of contacted hBN/BLG/hBN Hall bar before (left panel) and after (right panel) the top gate deposition (scale bar is 4~$\mu$m).}
	\label{transfer}
\end{figure}
\begin{figure}
	\centering
	\includegraphics[width=0.97\linewidth]{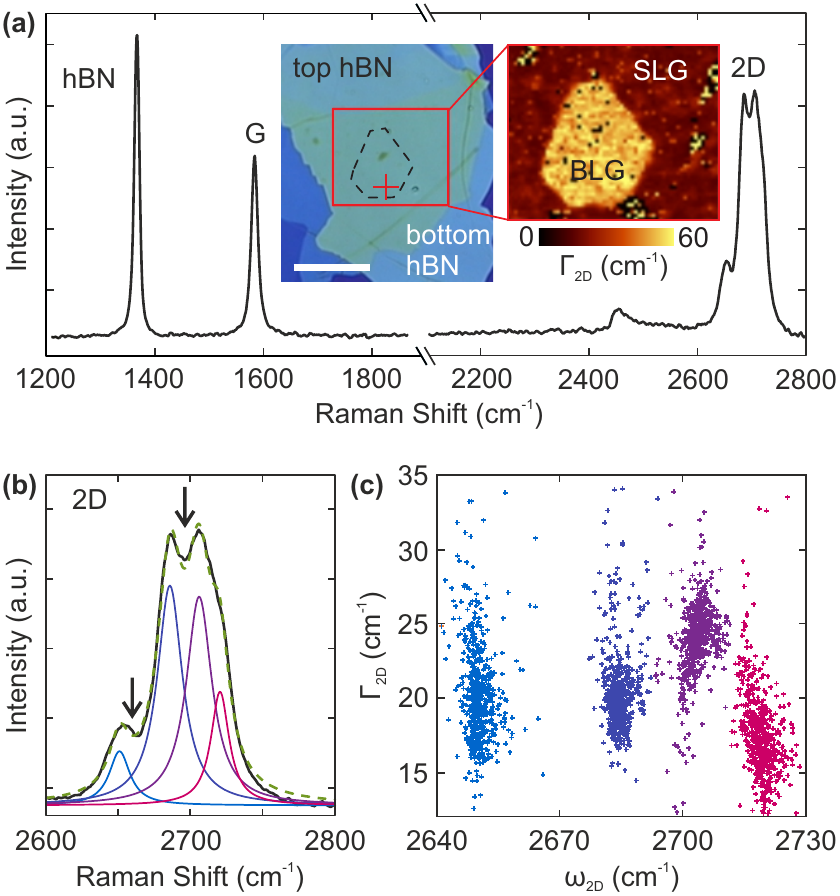}
	\caption{(a) Raman spectrum of a hBN/CVD-BLG/hBN heterostructure. The inset shows an optical image of the structure (scale bar: 20~$\mu$m). The black line outlines the BLG crystal. The BLG can easily be identified in the Raman map (right panel of inset) which shows the total 2D line width of both  BLG and SLG.  (b) Close-up of the 2D-peak in (a) with a fit curve (black dotted line) using four Lorentzians. (c) $\Gamma_\text{2D}$ of all 2D line sub-peaks vs. respective 2D-sub-peak positions $\omega_\text{2D}$ extracted from a Raman map recorded over the full BLG area in (a) with a color code equal to (b).}
	\label{raman}
\end{figure}

We use low pressure CVD growth of graphene on the inside of copper foil enclosures following the same recipe as reported by us for SLG.\cite{Banszerus2015,Li2011} Using this recipe, up to 10~\% of the graphene flakes exhibit aligned bilayer parts in their centers grown underneath the SLG crystals (see optical image in Fig.~\ref{transfer}(a)) with lateral sizes varying between 20 and 100~$\mu$m. After growth, the copper surface is left under ambient conditions for at least five days which results in an oxidation of the copper-to-graphene interface, weakening the respective adhesion force. We note that the oxidation process for BLG crystals is significantly slower than for SLG crystals.
For detaching of the graphene crystals we fabricate a transfer stamp by covering a glass slide with scotch tape and a layer of poly(vinyl-alcohol) (PVA) and poly(methyl-methacrylate) (PMMA). Subsequently, we exfoliate hBN crystals onto this stack. The PVA/PMMA is cut around a desired flake, detached from the glass slide and rested on a poly(dimethylsiloxane) (PDMS) block.  Using a mask aligner, the hBN/PMMA/PVA/PDMS stamp is brought into contact with the CVD-grown BLG and heated for one minute at 125~$^\circ$C. A schematic of the pick-up process is illustrated in Fig.~\ref{transfer}(b). After mechanical pick up of the BLG from the copper foil the hBN/BLG stack is placed on another hBN flake exfoliated on Si/SiO$_2$ which was cleaned beforehand by an oxygen plasma (base pressure of 0.1~mbar, 600~W for 300~s) and the PDMS block is thereafter removed. The resulting hBN/BLG/hBN heterostructure on SiO$_2$ is cleaned by in DI water acetone and isopropanol. The inset in Fig.~\ref{raman}(a) shows an optical false color image of a typical hBN/graphene/hBN heterostructure where the bilayer part is highlighted by a black dashed line.


We first characterize the hBN/BLG/hBN heterostructures by confocal Raman spectroscopy using a laser excitation wavelength of 532~nm. Fig.~\ref{raman}(a) shows a typical Raman spectrum detected on one of our structures recorded at the cross as displayed in the optical image of the inset. The Raman peak originating from the hBN is at $\omega_\text{hBN}$~=~1366~cm$^{-1}$.\cite{SMLL:SMLL201001628} Moreover, the graphene G-peak is located at $\omega_\text{G}$~=~1583~cm$^{-1}$ and exhibits a full-width-at-half-maximum (FWHM) of $\Gamma_\text{G}$~=~ 13~cm$^{-1}$ indicating an low overall doping.\cite{Yan2007, Pisana2007} The graphene 2D-peak at around 2700~cm$^{-1}$ exhibits the well-known BLG sub-peak structure. \cite{Graf2007} Fig.~\ref{raman}(b) shows a close-up of the 2D-peak where four Lorentzians are needed to fit the spectrum which proves the AB-stacking of our CVD-BLG.\cite{Herziger2014} Furthermore, as a result of the low FWHM of the individual sub-peaks, we observe distinct minima between the left peaks (see arrows in Fig.~\ref{raman}(b)), which suggests low nanometer-scale strain variations in our CVD BLG.\cite{Neumann2015, Couto2014, Engels2014a,Banszerus2017}.

To gain better statistical information on the strain variations over the BLG crystal, we record a high resolution Raman map in the area marked by the red box in Fig.~\ref{raman}(a). For each spectrum we fit both the peak position and the peak width of all four sub peaks of the 2D-peak. The result is summarized in a scatter plot in Fig.~\ref{raman}(c) where we plot respective $\Gamma_\text{2D}$ values for each sub-peak as function of their peak positions. The obtained distributions are in good agreement with data reported for high-quality exfoliated BLG\cite{Engels2014a}. Together with the spatial homogeneity seen in the Raman map of the 2D line width (inset of Fig.~\ref{raman}(a)) this underlines the high structural quality of the sample.

\begin{figure}
	\centering
	\includegraphics[width=0.97\linewidth]{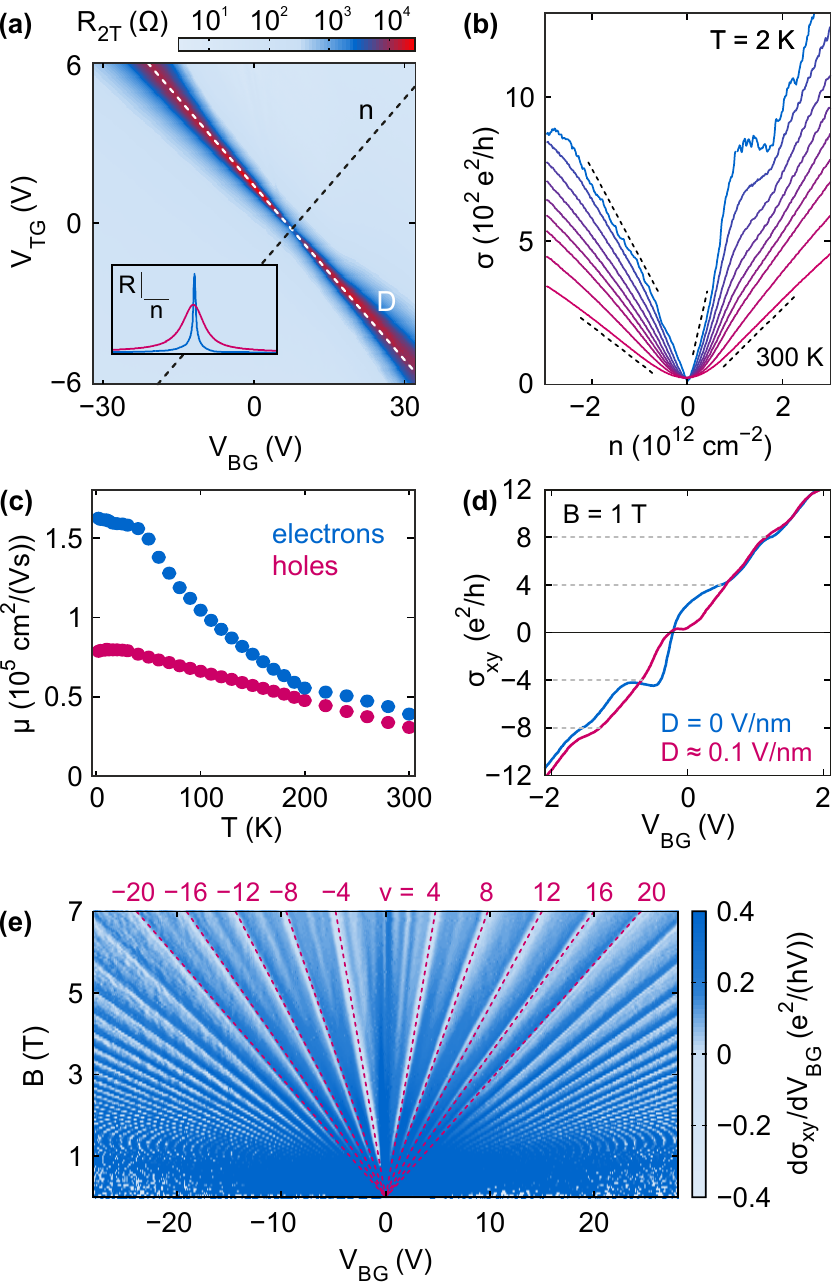}
	\caption{(a) Two terminal resistance R$_\text{2T}$ of the CVD BLG device in Fig.~1(d) vs. $V_\text{TG}$ and $V_\text{BG}$ (T=2~K). The dashed lines represent axes with only displacement field $D$ (white) and charge carrier density $n$ (black) variations. The inset shows the four terminal resistance $R$ (scale bar: 1~k$\Omega$) vs. $n$ (scale bar: $10^{12}$~cm$^{-2}$) for $D=0$~V/nm at 2~K (blue) and at 300~K (purple).  (b) $\sigma$ vs. $n$ for T~=~2, 40, 70, 100, 130, 160, 190, 240 and 300~K. Black dashed lines indicate linear fits to $\sigma$ for 2 and 300~K. The plateau on the electron side marks the onset of ballistic transport. (c) Electron (blue) and hole (purple) mobilities $\mu$ vs. $T$. (d) $\sigma_\text{xy}$  vs. $V_\text{BG}$ taken in a perpendicular magnetic field of $B$~=~1~T for D~=~0 (blue) and D~=~0.1~V/nm (purple). (e) $\text{d}\sigma_\text{xy}/\text{d}V_\text{BG}$ vs. $V_\text{BG}$ for different $B$. Landau levels with filling factors $\nu$~=~-20,..,20 are highlighted by dashed lines.}
	\label{transport}
\end{figure}

For magneto-transport experiments, we pattern the hBN/BLG/hBN heterostructure into a Hall bar using standard electron beam lithography, reactive ion etching and metal evaporation. The latter is used to fabricate one-dimensional side contacts (see Ref.\citenum{Banszerus2015} for details). The left panel of Fig.~\ref{transfer}(c) shows a contacted Hall bar prior to the deposition of the top gate. For fabricating the top gate, another hBN flake is transferred onto the device and a Cr/Au top gate is deposited over the entire structure. A finished device is shown in the lower panel of Fig.~\ref{transfer}(c). We note that the top gate exceeds the outer edges of the Hall bar transport channel to suppress inhomogeneous gate tuning along the edges of the BLG.

To characterize the tuning behavior of the Hall bar structure by means of low temperature transport measurements we record the two-terminal resistance $R_\text{2T}$ as a function of both top gate voltage ($V_\text{TG}$) and back gate voltage ($V_\text{BG}$) where the latter is applied across a 285 nm thick SiO$_2$ layer to the underlying highly doped Si wafer. The resulting two-dimensional map is shown in Fig.~\ref{transport}(a). A line of elevated resistance (white dotted line) can be tuned by both $V_\text{TG}$ and $V_\text{BG}$. Along this line the vertical displacement field $D$ is tuned at $n=0$~cm$^{-2}$. Every perpendicular path will change $n$ at $D=\text{const.}$ In a simple plate capacitor model, the displacement field equals to
\begin{equation}
D  = \frac{e}{2\epsilon_0}\left[\alpha_\text{BG}(V_\text{BG}-V_\text{BG,0}) - \alpha_\text{TG}(V_\text{TG}-V_\text{TG,0})\right]
\end{equation}
with the charge carrier density being
\begin{equation}
n  = \alpha_\text{BG}(V_\text{BG}-V_\text{BG,0}) + \alpha_\text{TG}(V_\text{TG}-V_\text{TG,0}),
\end{equation}
where $\alpha_\text{BG}$~=~6.65$\times10^{10}$~cm$^{-2}$V$^{-1}$ and $\alpha_\text{TG}$~= 31.37$\times10^{10}$~cm$^{-2}$V$^{-1}$ are the respective gate lever arms which are extracted from independent measurements of the Hall effect and the corresponding offset voltages $V_\text{BG,0}$~=~7.500~V and $V_\text{TG,0}$~=~-0.2212~V result from the finite doping of the sample. Using the lever arm we estimate the total thickness of the top gate dielectrics to be $\approx$~68~nm. The device shows an on-off ratio of more than 10,000 which is comparable to that achieved for exfoliated BLG.\cite{Yan2012b}

We now focus on the transport properties at zero displacement field $D=0$~V/nm. The inset of Fig.~\ref{transport}(a) shows the four-terminal BLG resistance $R$ measured between contacts A and B (see Fig.~1(d)) as function of $n$ at the two temperatures of 2 (blue) and 300~K (purple), where $V_\text{BG}$ and $V_\text{TG}$ are simultaneously varied along the red dashed line in Fig.~\ref{transport}(a). The maximum resistance is reached at the charge neutrality point (CNP).\cite{CastroNeto2009} Fig.~\ref{transport}(b) shows the corresponding conductivity
 $\sigma = l/(w R)$
for a larger set of temperatures between 2 and 300~K, where $l=$~2~$\mu$m is the length between the probing contacts A and B and $w=$~1~$\mu$m is the width of the Hall bar. From the $n$-dependence of $\sigma$ we extract the charge carrier mobilities by performing fits to the Drude formula $\sigma = e n\mu$. At 2~K, we find $\mu = 180,000$~cm$^2$/(Vs) for electrons and $\mu = 80,000$~cm$^2$/(Vs) for holes. The respective values at 300~K are $40,000$~cm$^2$/(Vs) and $30,000$~cm$^2$/(Vs). Fits are included as black dashed lines in Fig.~\ref{transport}(b), which are vertically offset for clarity \cite{BLGnote}. On the electron side ($n>0$) there is a plateau evolving near $n_\text{p}$~=~10$\time10^{12}$cm$^{-2}$ for temperatures below 200~K which is fully developed at 2~K. We attribute this feature to a transition from diffusive to ballistic transport. To support this notion, we consider the electron mean free path $l_\text{m} = (h/2e)\mu(n/\pi)^{1/2}$ with $h$ being the Planck constant, which has to exceed the distance between the voltage probes
 $l=$~2~$\mu$m in order to allow for ballistic transport. Together with $n_\text{p}$ this provides an estimate for the electron mobility of $\mu = 175,000$~cm$^2$/(Vs) for the device to become ballistic which is in good agreement to the extracted low temperature values. The broadening of the plateau at elevated temperature marks the crossover from ballistic to diffusive transport due to temperature-induced electron-phonon scattering. A detailed evaluation of the temperature dependent electron and hole carrier mobilities is shown in Fig.~\ref{transport}(c) where we restrict our fitting range to the diffusive transport regime at smaller carrier densities for $T~<~$200~K, where $l_\text{m}<l$. Remarkably, the room temperature mobilities are about two times larger than previously reported values for CVD BLG~\cite{Hao2016} and match those values obtained in state-of-the-art heterostructures fabricated from exfoliated BLG.\cite{Dean2010,Engels2014,Zhu2017}

We now explore the quantum Hall effect (QHE) in our hBN/BLG/hBN Hall bar device. Fig.~\ref{transport}(d) shows the transverse conductance $\sigma_\text{xy}$ as function of $V_\text{BG}$ measured in a perpendicular magnetic field of $B$~=~1~T for two displacement fields. The measurements show a clear Landau level formation with steps of 4~e$^2$/h. These steps together with the observation that an additional Landau level emerging at 0~e$^2/$h with non-zero displacement field directly proves the AB stacking order in our CVD-BLG.\cite{Novoselov2006} To further visualize the magnetic field dependence of the QHE, we plot in Fig.~\ref{transport}(e) the derivative of $\sigma_\text{xy}$ with respect to $V_\text{BG}$ as function of $V_\text{BG}$ and $B$. Landau level formation is  seen above $B$~=~1~T with filling factors $\nu$ being multiples of four. Furthermore, there is a partial degeneracy lifting at larger $B$-fields. This observation is in good agreement with measurements performed on high quality exfoliated BLG\cite{Engels2014, Novoselov2006} and underlines the remarkable electronic quality of our CVD-grown BLG.

 \begin{figure}
	\centering
	\includegraphics[width=0.97\linewidth]{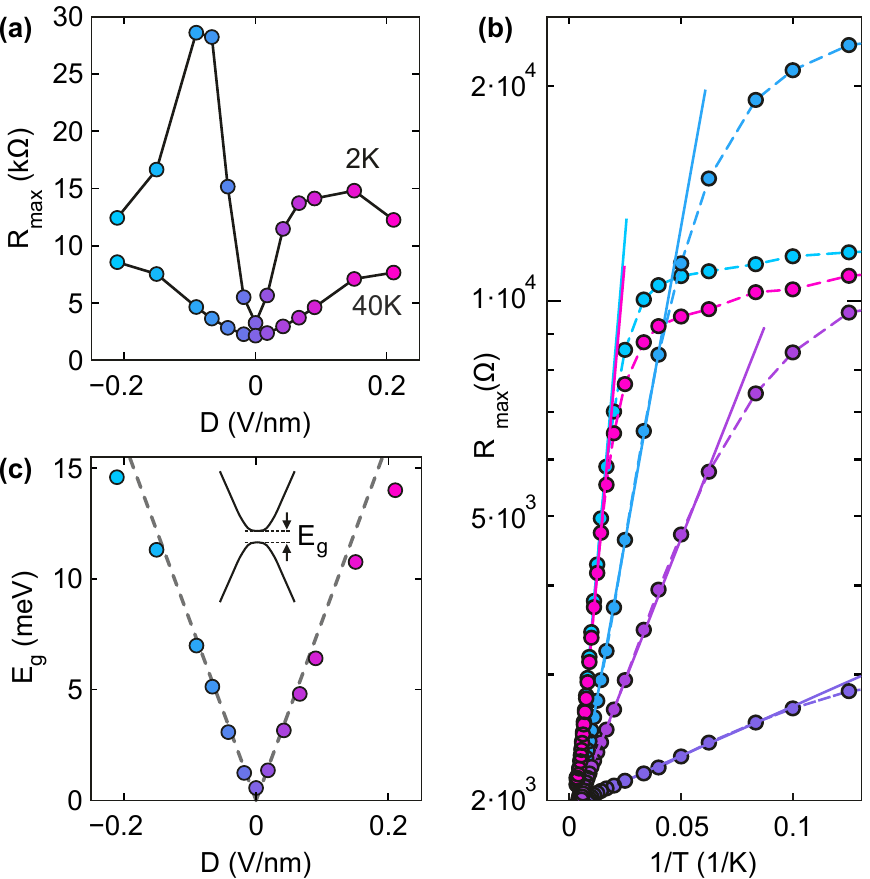}
	\caption{(a) Two terminal maximum resistance $R_\text{max}$ at the CNP ($n=0$) as function of displacement field $D$ for T~=~2~K and T~=~40~K. (b) $R_\text{max}$ as function of the inverse temperature $T^{-1}$ for various $D$ values according to the colors in (a). The solid lines are fits to a temperature activated transport at higher temperatures. (c) Extracted activation energies (energy gaps) $E_\text{g}$ vs. $D$ which are compared to theoretical predictions (gray lines).\cite{Zhang2008, Zhang2009}}
	\label{dual}
\end{figure}

In the following, we focus on the displacement field-induced formation of the energy band gap. Fig.~\ref{dual}(a) shows the maximum
two-terminal resistance $R_\text{max}$ measured at the CNP along the white dashed line in Fig.~\ref{transport}(a) as a function of the displacement field $D$. $R_\text{max}$ exhibits a global minimum at $D=0$~V/nm and continuously increases for both positive and negative $D$-fields as expected for band gap formation. At low temperature ($T=2$~K), $R_\text{max}$ shows maximum values at around $|D| = 0.1$~V/nm and decreases again at larger displacement fields. Such a maximum is not observed at 40~K. The low temperature behavior contradicts to ideal AB-stacked BLG where $R_\text{max}$ is expected to increase monotonically with increasing $|D|$-values, but has also been observed in devices with exfoliated BLG.\cite{Zhang2008, Zhang2009, McCann2006b, Min2007,Zhu2017}

In order to analyse the low temperature transport behavior, we plot in Fig.~\ref{dual}(b) $R_\text{max}$ as a function of $1/T$ on a semi-log scale.  As expected for thermally activated transport across an energy gap, $R_\text{max}$ varies linearly at higher temperatures (low $1/T$ values). The energy gap $E_\text{g}$ can be extracted from\cite{Taychatanapat2010}
\begin{align}
  R_\text{max} (T) \propto \exp\left(\frac{E_\text{g}}{2k_\text{B}T}\right),
 \label{eq_transport}
\end{align}
where $k_\text{B}$ is the Boltzmann constant. The corresponding fits for selected $D$ fields (color code identical to the data in Fig.~\ref{dual}(b)) show good agreement for low $1/T$ values. In Fig.~\ref{dual}(c) we plot the extracted $E_\text{g}$ as function of $D$. The values of $E_\text{g}$ exhibit a distinct minimum at $D=0$~V/nm and show a monotonic and symmetric increase towards positive and negative $D$ fields. The gray lines in Fig.~\ref{dual}(c) represent theoretical predictions for BLG\cite{Zhang2008, Zhang2009} highlighting the good agreement with the expected behavior for the band gap opening in AB-stacked BLG. We note that the observed dependence of $E_\text{g}$ on $D$ clearly indicates the opening of a band gap of up to 15~meV while the deviation from the thermally activated behavior at low temperatures (see Fig.~\ref{dual}(b)) points to the presence of additional transport channels. These channels may result from the presence of stacking folds which have been previously reported in BLG~\cite{Alden2013,SanJose2014,Yin2016}. They can lead to the formation of AB/BA stacking domain walls which may serve as ballistic transport channels~\cite{Ju2015} at sub-gap energies. Another possible explanation could be the formation of conducting edge channels.\cite{Zhu2017} The conductance in our device at the CNP is on the order of few e$^2$/h which matches expectations for ballistic transport channels shunting the device. As our device becomes increasingly conductive with increasing temperature, the contribution of one-dimensional channels strongly weakens due to their metallic behavior.\cite{Ju2015} Further investigations are needed to explore these transport channels in more detail.

In summary, we applied a dry-transfer pickup technique to delaminate CVD-grown BLG from the catalytic copper substrate using exfoliated hBN crystals. Raman spectroscopy and quantum Hall effect measurements prove an AB stacking order and a high electronic quality of our BLG crystals. Transport measurements on hBN/CVD-BLG/hBN Hall bar structures reveal very high charge carrier mobilities of up to 40,000~cm$^2$/Vs at 300~K which is about two times larger than previously reported values for CVD based BLG devices. We demonstrate the opening of a band gap of up to 15 meV when applying vertical displacement field of 0.2 V/nm. While we have identified sub-gap transport channels at the lowest temperatures which might be attributed to AB/BA stacking folds we demonstrate a device performance which well compares to state-of-the-art exfoliated BLG-based devices.

Work supported by the EU project Graphene Flagship (contract no. 696656) and the DFG (BE 2441/9-1). Growth of hexagonal boron nitride crystals was supported by the Elemental Strategy Initiative conducted by the MEXT, Japan and JSPS KAKENHI Grant Numbers JP26248061, JP15K21722 and JP25106006.


\begin{thebibliography}{42}
\bibitem [{BLG()}]{BLGnote}%
  \BibitemOpen
  \href@noop {} {}\bibinfo {note} {While we used the same fitting range at all
  temperatures for the hole side ($n<0$), there is ballistic transport evolving
  at lower temperatures above $n=10^{12}$cm$^{-2}$ on the electron side. We
  therefore have to restrict the Drude fit, which only applies for diffusive
  transport, to lower carrier densities below 160 K where the $\sigma$ vs. $n$
  dependence also becomes linear towards lower $n$ values (black dotted line
  for $n > 0$ at $T = 2$~K). On the hole side, the $\sigma$ vs. $n$ slope also
  becomes steeper towards the CNP at lower temperatures. The extracted hole
  mobilities are thus conservative estimates at lower temperatures. For
  example, if we use the same fitting region as for the electron transport at
  2~K we extract hole mobilites of 100,000 cm$^2$/(Vs).}\BibitemShut {Stop}%
\end{thebibliography}
\end{document}